# On the screening of the potential between adjoint sources in $QCD_3$

Grigorios I. Poulis[a] * † and Howard D. Trottier[b] ‡

[a] NIKHEF-K, Theory Group, Postbus 41882, 1009 DB Amsterdam, The Netherlands

[b]Physics Department, Simon Fraser University, Burnaby, British Columbia V5A 1S6, Canada

We calculate the potential between adjoint sources in $SU(2)$ pure gauge theory in three dimensions. We investigate whether the potential saturates at large separations due to the creation of a pair of gluelumps, colour–singlet states formed when glue binds to an adjoint source.

## 1. INTRODUCTION

The formation of chromoelectric flux tubes between static quarks is by now a well established feature of lattice QCD simulations [1,2]. Of particular interest to the understanding of the mechanisms of both hadronization and confinement are studies of the conditions under which such flux tubes break. In zero temperature quenced QCD simulations the absence of dynamical fermions does not allow the screening of the potential by virtual colour–singlet $q\bar{q}$ pairs and thus the interquark potential is expected to rise linearly with the separation $R$ for arbitrarily large $R$. However, if one considers a pair of adjoint (as opposed to fundamental) representation sources then one expects complete screening at large $R$ where a pair of colour–singlet states ("gluelumps" [3]) are formed by coupling the adjoint "quarks" to (adjoint) glue. One naively expects that the adjoint potential will saturate to an $R$–independent value for distances $R$ such that [3]

$$\sigma_{adjoint} \cdot R \geq 2 \cdot M_{gluelump} \ , \tag{1}$$

where $\sigma_{adjoint}$ is the adjoint string tension and $M_{gluelump}$ is the mass of the lightest gluelump. This expectation has been tested in pure gauge $SU(2)$ in four dimensions by Michael [3,4]. Although the gluelump mass was measured quite accurately, the data on the adjoint potential do not allow one to conclude whether the adjoint potential saturates.

To shed some light into this issue we present here results on the three dimensional pure gauge $SU(2)$ theory ($QCD_3$), which, like $QCD_4$, is a confining theory with a nontrivial glueball spectrum [5]. The first calculation of the gluelump mass in $QCD_3$ is presented here. We also compute the adjoint $Q\bar{Q}$ potential to much larger $R$ than previously available [6] by using fuzzing techniques [7]. Our results provide for the first test of the naive screening hypothesis, Eq. (1), in $QCD_3$.

## 2. METHOD

### 2.1. Observables

In order to compute the potential in different representations of $SU(2)$, we measure temporal $R \times T$ Wilson loops constructed from fundamental links and then obtain the corresponding Wilson loops in higher representations, $W_I(R,T)$, (where for $SU(2)$ the fundamental representation corresponds to $I = 1/2$, the adjoint to $I = 1$, quartet to $I = 3/2$, etc.) using relations amongst the group characters (see for example [2]). For the calculation of the gluelump mass we construct the "dumbbell" operator [3]

$$\begin{aligned} \mathcal{O}(T) &= \mathrm{Tr}\big(F_0\sigma^a\big)\ A_{ab}(Q)\ \mathrm{Tr}\big(F_T^\dagger\sigma^b\big) \qquad (2) \\ &= 2\Big[\mathrm{Tr}\Big(F_0 Q F_T^\dagger Q^\dagger\Big) - \frac{1}{2}\mathrm{Tr}(F_0)\mathrm{Tr}(F_T)\Big] \ , \end{aligned}$$

involving the time propagation of a static adjoint quark $A(Q)$ coupled at time 0 and time

*presented by

†supported by Human Capital & Mobility EC Fellowship ERBCHBICT941430

‡supported in part by the Natural Sciences and Engineering Research Council of Canada

$T$ to a gluon fields (spatial plaquettes) $F_0$ and $F_T$, respectively (see Figure 1). Here $A^{ab}(Q) \sim \mathrm{Tr}(\sigma^a Q \sigma^b Q^\dagger)$ is the adjoint representation of the product of the timelike fundamental links $Q = U_3(0) \ldots U_3(T)$, and $\{\sigma^a\}$ are Pauli matrices.

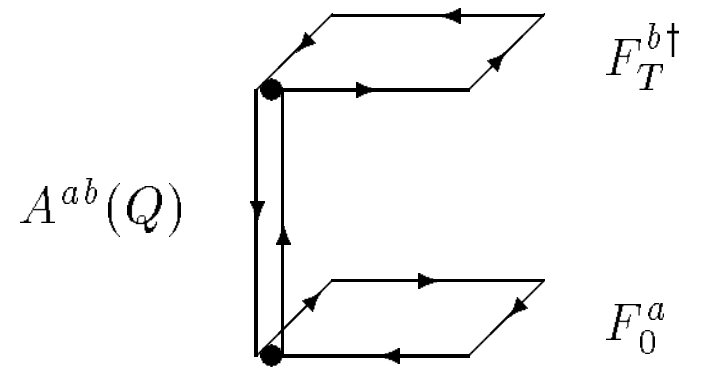

Figure 1. The gluelump operator.

For $SU(2)$ we see that from the two combinations $F \pm F^\dagger$ only the "$-$" contributes [3]; hence only negative charge–parity gluelumps exist. Here we present results for the "magnetic" gluelump involving the combination of spatial plaquettes around the $t = 0$ or $t = T$ ends of the adjoint timelike link product depicted in Figure 2. The lowest angular momentum excited by this operator is (in three dimensions) $J = 0$. We then obtain the potential and the mass as the $T \to \infty$ extrapolation of effective (i.e. $T$–dependent) quantities

$$\begin{aligned} V_I^{eff}(R,T) &= \ln\Big[W_I(R,T)/W_I(R,T+1)\Big] \\ M^{eff}(T) &= \ln\Big[\mathcal{O}(T)/\mathcal{O}(T+1)\Big] \ . \end{aligned} \quad (3)$$

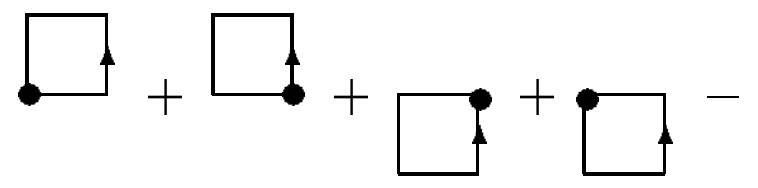

Figure 2. Combination of spatial plaquettes for the magnetic gluelump.

### 2.2. Techniques

For the *spacelike* links $U_\mu(x)$ we employ an iterative fuzzing scheme, and obtain new links

$$U_\mu(x) \to \mathcal{P}\Big\{\mathcal{C} \cdot U_\mu(x) + \sum_{\nu \neq \mu} \text{U-staples}_{\mu,\nu}\Big\} \ , \quad (4)$$

where $\mathcal{C}$ is a positive constant, $\mathcal{P}$ is a multiplicative projector onto $SU(2)$ and this procedure is iterated $\mathcal{N}$ times. $\mathcal{C}, \mathcal{N}$ are chosen so as to maximize the overlap with the ground state without too much computer time cost. On the *timelike* links we apply an analytic version of the first order multihit procedure of Parisi et. al. [8]. We use the standard Wilson action with periodic boundary conditions and with a heatbath update. We thermalize between 6000 (at $\beta = 6$) and 10000 times (at $\beta = 12$) and take measurements separated by 20 to 40 updates.

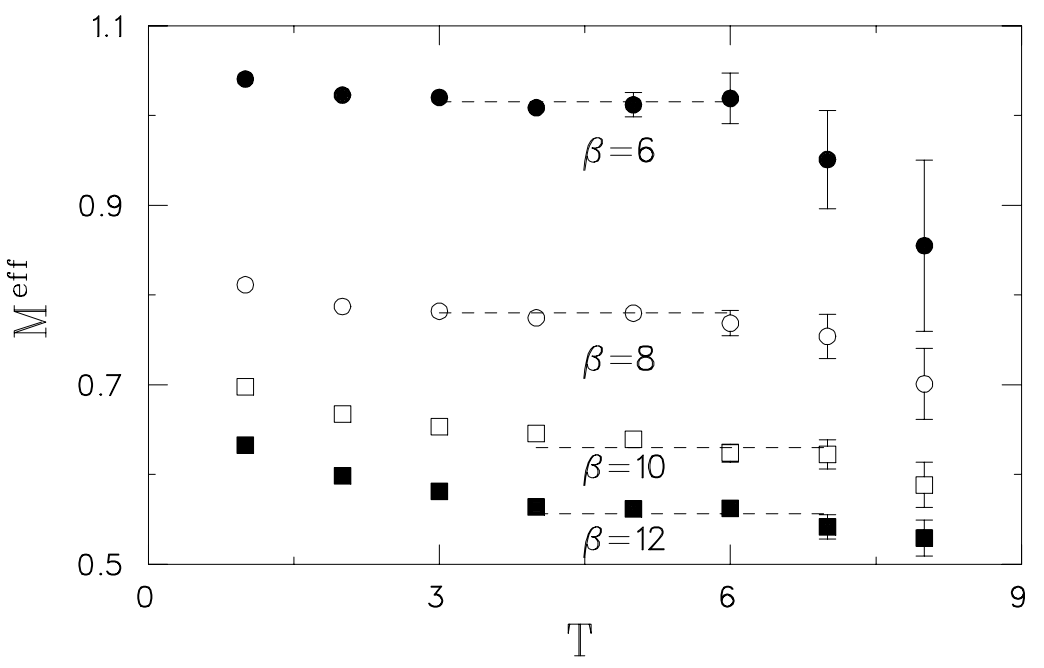

Figure 3. Effective magnetic gluelump mass at various $\beta$ values.

## 3. RESULTS

### 3.1. Magnetic Gluelump

For the calculation of the magnetic gluelump mass we compute the expectation value of the $\mathcal{O}(T)$ operator, Eq. (2), for $T \in \{1,8\}$ from 2000 measurements on a $24^3$ lattice for $\beta = 6, 8, 10$ and 12, with fuzzing parameters $(\mathcal{C}, \mathcal{N}) = (2,7)$. The results are presented in Figure 3; following Ref. [4] we identify the first $T, T+1$ pair where the effective values are equal within statistical errors (this is $T = 4$ for the gluelump and $T = 3$ for the adjoint potential) and take $M^{eff}(T)$ with its statistical error as an estimate of the true mass $M^{eff}(T \to \infty)$. Since any effective value defines an upper limit for the true ground state a crude estimate of the systematic errors for this extrapolation procedure may be given by the statistical error of the $T+1$ value. Thus, we find that at $\beta = 6$ the magnetic gluelump mass (including self energy) is $1.009 \pm 7(\text{stat.}) - 16(\text{syst.})$. That means that at $\beta = 6$ the adjoint potential should saturate to about 2, if the screening by gluelumps hypothesis is correct. We find that in physical units the gluelump mass is about $6/\beta$ at $\beta = 6$ and about $6.5/\beta$ at $\beta = 12$, thus it scales at the 10% level.

### 3.2. Adjoint Potential

In order to check the breaking of the adjoint string we try to probe the largest physical length available (i.e. low $\beta$ for fixed lattice size) while remaining inside the scaling region; thus we present results for the potential at $\beta = 6$. For the calculation of the potential we compute timelike $(R \times T)$ Wilson loops with $R \in \{3, 16\}$ and $T \in \{1, 5\}$. Our main results come from 4000 measurements on a $32^3$ lattice and 5000 measurements on a $40^3$ lattice with fuzzing parameters $(\mathcal{C}, \mathcal{N}) = (2.5, 10)$.

Table 1
Effective adjoint potential $V_1^{eff}(R,T)$ at $\beta = 6$

| $T$ | $R = 10$ | $R = 11$ | $R = 12$ |
|---|---|---|---|
| 1 | 2.240(1) | 2.425(1) | 2.610(1) |
| 2 | 2.157(2) | 2.330(2) | 2.503(3) |
| 3 | 2.124(6) | 2.291(10) | 2.462(16) |
| 4 | 2.139(28) | 2.326(57) | 2.507(116) |
| 5 | 2.096(136) | 2.263(329) | 2.645(979) |

Table 2
The adjoint potential $V_1(R)$ at $\beta = 6$. First error is statistical, second systematic (see text)

| $R$ | Ref. [6] | this work | $R$ | this work |
|---|---|---|---|---|
| 3 | 0.910(4) | 0.912(0)(0) | 10 | 2.124(6)(28) |
| 4 | 1.105(3) | 1.103(0)(1) | 11 | 2.291(10)(57) |
| 5 | 1.280(11) | 1.281(1)(1) | 12 | 2.462(16)(116) |
| 6 | 1.478(20) | 1.455(1)(3) | 13 | 2.631(27)(209) |
| 7 | 1.644(43) | 1.625(2)(6) | 14 | 2.792(45)(360) |
| 8 | | 1.793(3)(11) | 15 | 2.935(74((435) |
| 9 | | 1.964(6)(22) | 16 | 3.017(113)(708) |

From Table 1 we see that the $T = 3$ and $T = 4$ values of $V_1^{eff}$ agree within statistical errors. We test our procedure for assigning values for the potential from the $T = 3$ slice by comparing (with excellent agreement) our values for $R = 3 - 7$ with the published values of Mawhinney [6] in Table 2. In Figure 4 we plot the adjoint and the fundamental potential (the latter scaled by the ratio of Casimirs $C_1/C_{1/2}$, $C_I \equiv I(I+1)$) as a function of the separation. The results are somehow ambiguous due to the large systematic errors (lower error bars). However, there is some indication that a much larger separation $R$ is required to observe significant screening than is suggested by the naive screening hypothesis of Eq. (1). We are currently accumulating more data in order to reduce the systematic errors which will allow for a more definitive assessment of the naive screening hypothesis.

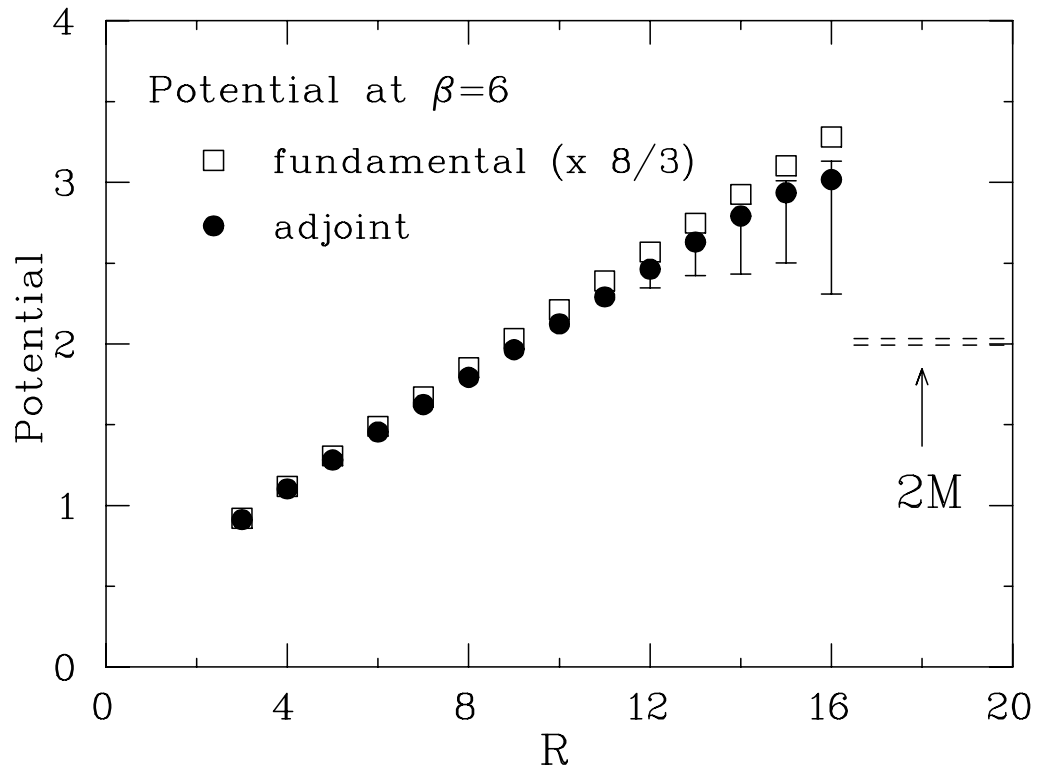

Figure 4. Potential at $\beta = 6$. The value of twice the magnetic gluelump mass is also depicted.

## 4. ACKNOWLEDGEMENTS

We thank R. Woloshyn for fruitful discussions.